\documentclass{elsarticle}

\usepackage{paradise,longtable}

\newcommand{\hypertarget}[2]{}
\newcommand{\href}[2]{#2}

\begin{document}

\begin{frontmatter}

\title{\divm{}: Model Checking with \llvm{} and Graph Memory\tnoteref{t1}}
\tnotetext[t1]{This work has been partially supported by the Czech Science Foundation
grant No. 15-08772S and by Red Hat, Inc.}
\author{Petr Ročkai}  \ead{xrockai@fi.muni.cz}

\author{Vladimír Štill}  \ead{xstill@fi.muni.cz}

\author{Ivana Černá}  \ead{cerna@fi.muni.cz}

\author{Jiří Barnat}  \ead{barnat@fi.muni.cz}

\address{\fimuni}

\begin{abstract}
  In this paper, we introduce the concept of a virtual machine with
  graph-organised memory as a versatile backend for both explicit-state
  and abstraction-driven verification of software. Our virtual machine
  uses the \llvm{} IR as its instruction set, enriched with a small set of
  \emph{hypercalls}. We show that the provided hypercalls are sufficient
  to implement a small operating system, which can then be linked with
  applications to provide a POSIX-compatible verification environment.
  Finally, we demonstrate the viability of the approach through a
  comparison with a more traditionally-designed \llvm{} model checker.
\end{abstract}

\end{frontmatter}

\section{Introduction}\label{introduction}

Applying verification to real-world programs is undoubtedly desirable --
it can increase code quality while cutting costs at the same time. Model
checking is one of the approaches that can provide robust correctness
guarantees without introducing false positives. This precision, however,
does not come for free -- model checking, especially in the context of
software, is computationally very expensive. Nonetheless, as our
previous work shows~\citep{rockai15:techniq.memory}, a combination of
state space reduction techniques, compression and of a tailored approach
to test case construction makes model checking a genuinely useful
programming aid. For example, we have successfully applied this approach
in development of scalable concurrent data
structures~\citep{barnat15:fast.dynamic} in C++.

\subsection{Application Area}\label{application-area}

The main area we are aiming at in this paper is verification of C and
C++ programs that do not explicitly interact with their environment and
in particular, do not read data from uncontrolled outside sources. As an
example, it is permissible for the program to read from a fixed file, in
the understanding that the content of the file is treated as part of the
program, not as variable (arbitrary) input. Implicit interactions are,
however, allowed: thread scheduling is taken to be arbitrary, as are
various failure scenarios like an ``out of memory'' condition during a
\texttt{malloc} call.

Typically, computer programs are made from components of various size,
ranging from individual functions and classes, through units and
libraries to frameworks and complete applications. At the highest
levels, static analysis can process large amounts of code in bulk,
pointing out possible problems, with varying levels of precision. At the
lowest level, a small number of well-isolated, well-defined and highly
critical functions can be subjected to rigorous treatment via automated
theorem proving or exhaustive symbolic model checking. In this paper, we
are primarily concerned with the mid-low part of the component spectrum:
the unit level. Units are collections of coupled functionality and data
structures. It is common practice that individual units of a program are
tested separately, often by writing \emph{unit tests}: those are small,
self-contained programs that exercise the functionality provided by a
single unit (and indirectly also the functionality of its dependencies).

These \emph{unit tests} very often exactly reflect the constraints
outlined above: their interaction with the outside world is, by design,
very limited. However, implicit interactions -- thread scheduling,
memory pressure and similar effects -- are usually very hard to control
in a testing environment. This makes an explicit-state model checker,
which can defeat these remaining sources of non-determinism, an
extremely valuable tool.

Explicit-state model checking is, however, not the only application area
of the research presented in this paper; it is merely the primary one,
as it is the one that is best understood. Abstraction-driven and
symbolic approaches to software verification are a hot research topic,
and the contributions of this paper can be combined with advances in
those areas. We fully expect that such a combination will also work at
higher levels of abstractions: complete libraries and applications (see
also Section~\ref{sec:symbolic}).

\subsection{Goals}\label{sec:goals}

Our main goal is to design an abstract machine (\divm{}), that is, a
low-level programming language, with these two properties:

\begin{enumerate}
\def\labelenumi{\roman{enumi}.}
\item
  The machine should be a suitable target for compiling C and C++
  programs, including \emph{system-level} software (an operating system
  kernel and system libraries like \texttt{libc}).
\item
  An efficient implementation of the semantics of this abstract machine
  should be possible. It should be easy to store and compare states of
  the machine and to quickly compute the transition function.
\end{enumerate}

It is typical of contemporary software verification tools to include
\emph{ad-hoc} extensions of the C language. The reason for this is that
system-level software (per our first criterion) needs additional
facilities, not available in the C language. In other words, C as a
language is \emph{incomplete}: system-level software cannot be expressed
in the C language \emph{alone} (same is true of C++ and many other
languages). We would like to take, instead, a principled approach:
provide an abstract machine with sufficient expressive power.

\subsection{Contribution}\label{contribution}

We present an abstract machine (called \divm{}), based on the widely-used
\llvm{} IR\footnote{Intermediate Representation. The \llvm{} IR is used across
  the majority of the \llvm{} toolchain and is an abstract counterpart of
  the machine-level assembly language. Unlike machine-level languages,
  the \llvm{} IR is easy to transform and optimise automatically.} with a
small number of extensions. Both criteria outlined in
Section~\ref{sec:goals} are fulfilled by the proposed machine: it is
possible to express all the usual constructs -- such as threads,
processes, memory management and (simulated) input and output -- as
routines running on the machine, Moreover, since the machine is based on
\llvm{} IR, standards-compliant C and C++ compilers are readily available
targeting the machine. Additionally, we provide ports of crucial system
libraries: the C and C++ standard libraries, and a meaningful subset of
the POSIX interface. In particular, POSIX threads and POSIX file system
APIs are available.

Moreover, there are established methods for efficient implementation of
languages built around the concept of instructions which act on the
state of a machine. This is, after all, how computers operate on the
hardware level. Likewise, compilation of high-level, expression-based
languages with structured control flow (like C and C++) into low-level,
instruction-based languages (such as \llvm{}) is a well researched topic,
and high-quality implementations are available as off-the-shelf
components.

What we show in this paper is that addition of graph memory has no
detrimental effect on those established properties, and that, in fact,
it makes operations on the state of the machine more efficient.
Likewise, while the semantics are not, strictly speaking, simplified by
the addition of graph memory, it does make certain properties of the
program much easier to express. It is, therefore, our opinion, that the
addition of graph memory makes the machine and its semantics more
expressive in a meaningful way. The details of the graph structure of
the machine's memory are covered in Section~\ref{sec:graphmem}.

Finally, a reference implementation is available under a permissive,
open-source licence. All the source code relevant to this paper, along
with additional data and other supplementary material is available
online.\footnote{\url{https://divine.fi.muni.cz/2017/divm/}}

\subsection{Analysis and \llvm{}}\label{analysis-and-llvm}

\llvm{} is, primarily, a toolbox for writing compilers. Among other things,
this means that it is not a complete virtual machine, merely an
intermediate representation suitable for static analysis, optimisation
and native code generation. In particular, it may not always be possible
to encode an entire program in \llvm{} alone: compilers often work with
individual units, where undefined references are common and expected.
When the program is linked (whether statically or at runtime), these
unresolved references are bound to machine code, which may or may not be
derived from \llvm{} bitcode. A common example of such non-\llvm{}-derived
code would be the syscall interface of an operating system, which is
usually implemented directly in platform-specific assembly. At this
level, cooperation of code from various sources is facilitated by
machine-level calling conventions that live below the level of \llvm{}
bitcode.

An important consequence is that analyses that require complete
knowledge of the entire system cannot rely entirely on \llvm{} bitcode
alone. Different \llvm{}-based tools approach this problem differently. The
most common solution is to hard-wire knowledge about particular external
functions (i.e.~functions that usually come from system-specific
libraries that are not available in pure \llvm{} form, like
\texttt{pthread\_create} or \texttt{read}) into the tool. This \emph{ad
hoc} approach is suitable for experiments and prototypes, but is far
from scalable -- covering functionality commonly required by simple
programs entails hundreds of functions. To combat this problem, we
propose a small extension to the \llvm{} language, based on a small set of
\emph{hypercalls} (a list is provided in Table~\ref{tbl:hypercalls}).
Unlike pure \llvm{}, the \divm{} language is capable of encoding an operating
system, along with a syscall interface and all the usual functionality
included in system libraries.

\hypertarget{tbl:hypercalls}{}
\begin{longtable}[]{@{}ll@{}}
\caption{\label{tbl:hypercalls}A list of hypercalls provided by \divm{}.
}\tabularnewline
\toprule
Hypercall & Description\tabularnewline
\midrule
\endfirsthead
\toprule
Hypercall & Description\tabularnewline
\midrule
\endhead
\texttt{obj\_make} & Create a new object in the memory graph of the
program\tabularnewline
\texttt{obj\_free} & Explicitly destroys a object in the memory
graph\tabularnewline
\texttt{obj\_size} & Obtain the current size of an object\tabularnewline
\texttt{obj\_resize} & Efficiently resize an object
(optional)\tabularnewline
\texttt{obj\_shared} & Mark an object as \emph{shared} for τ reduction
(optional)\tabularnewline
\texttt{trace} & Attach a piece of data to an edge in the execution
graph\tabularnewline
\texttt{interrupt\_mem} & Marks a memory-access-related interrupt
point\tabularnewline
\texttt{interrupt\_cfl} & Marks a control-flow-related interrupt
point\tabularnewline
\texttt{choose} & Non-deterministic choice (a fork in the execution
graph)\tabularnewline
\texttt{control} & Read or manipulate machine control
registers\tabularnewline
\bottomrule
\end{longtable}

\subsection{Explicit-State Model
Checking}\label{explicit-state-model-checking}

Past experience has repeatedly shown that a successful explicit-state
model checker needs to combine a fast evaluator (the component which
computes successor states), partial order~\citep{peled98:ten.years}
and/or symmetry reductions~\citep{clarke98:symmet.reduct} and efficient
means to store the visited and open
sets~\citep{barnat10:scalab.shared, laarman14:scalab.multi}. The virtual
machine (VM) we propose covers the evaluator, but it also crucially
interacts with the remaining parts of the model checker.

\begin{figure}
\centering
\includegraphics{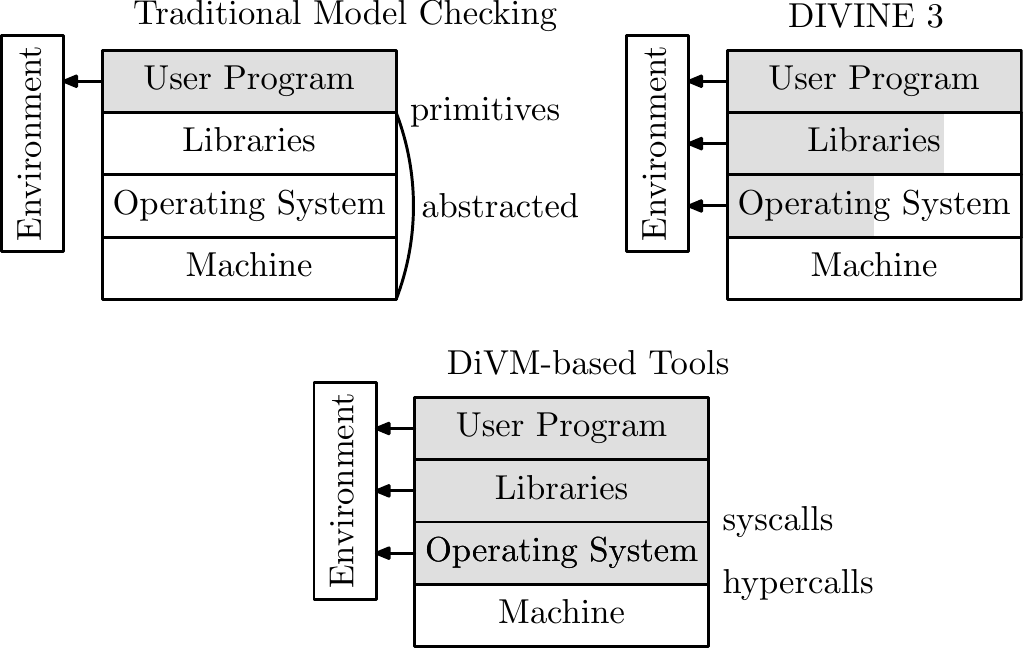}
\caption{Evolution of model checking. White boxes are built into the
model checker itself, shaded areas are part of the model (partially
supplied by the user, partially by the tool).}\label{fig:evolution}
\end{figure}

Moreover, as outlined above, the verifier also interacts with the system
under test (SUT). For our purposes, the SUT is not the user program
alone, but it also includes all libraries it is linked to (including
system libraries like \texttt{libc}) \emph{and} an operating system
(OS). This OS is, at least to some degree, bound to the VM it is
executing in and relies on its particular capabilities. In the case of
\divm{}, this includes the interfaces related to verification, i.e.~the
hypercall interface. To a lesser degree, these may also be used by
libraries which are part of the OS (typically \texttt{libc} and related
low-level code, eg. a thread support library). Overall, while the OS
itself is not very portable (running it on a typical hardware platform
would require extensive changes), it can host programs which work on
other systems, often without any modifications to the program.

From the semantic point of view, the VM comprises an abstract machine,
and its semantics should be such that it is possible to (sufficiently
faithfully) map C semantics\footnote{Or, to be more precise, the
  semantics of a C program executing in an operating system which
  provides additional facilities, like memory management.} onto the
semantics of the VM. The abstract machine executes a program, which is
composed of functions (routines), which are composed of instructions --
in our case, an instruction is either from the \llvm{} instruction
set~\citep{llvm16:llvm.languag} or it is a hypercall invocation.
Instructions manipulate the \emph{state} of the abstract machine: under
our proposed scheme, the state consists of two parts, a small, fixed set
of \emph{control registers} and of graph-structured memory (the
\emph{heap}).\footnote{The state is made available to the verifier via
  an interface of the virtual machine. The verifier is free to modify
  the state as needed, in particular, it can easily store the state
  (say, in a hash table) and reset the VM to that particular state
  later.} The nodes of the memory graph -- \emph{heap objects} -- are
byte arrays. Whenever a numeric representation of a \emph{pointer} is
stored in a node (at an arbitrary offset), an edge is created in the
graph, directed towards the heap object designated by the numeric
pointer. A set of \emph{root pointers} is stored in the control
registers: only objects reachable from this set are included in the
state. The semantics are then, of course, given by a function which
assigns, to a state and an instruction, a new state.

Our focus in this paper is twofold: first, design the hypercall
interface of the VM -- that is, describe the semantics of the abstract
machine which is realised by the VM and how they relate to semantics of
C programs; second, analyse the consequences of the chosen interface on
the various components of the verifier and the SUT.

\subsection{Design Motivation}\label{design-motivation}

There are three main advantages in the proposed approach to separation
of components. First, components adhering to a small and well-defined
interface can be much more easily re-used: for example, the OS part can
be re-used with a different model checker, saving a substantial amount
of work and thus reducing the barrier for achieving practicality for new
tools. The \texttt{pthread} library alone comprises more than 100
functions which can be shared by multiple tools as long as they expose
the same 10 basic hypercalls described in this paper. The same applies
to hundreds of syscalls available in a modern Unix-like OS.

\begin{figure}
\centering
\includegraphics{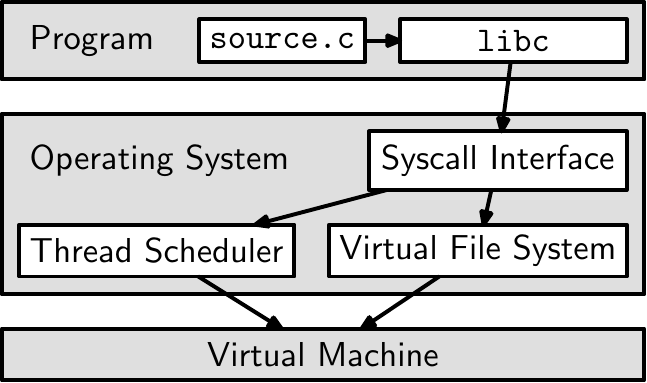}
\caption{Scheme of the execution and/or verification
environment.}\label{fig:scheme}
\end{figure}

Secondly, a small interface between the VM and the OS makes it quite
easy to write special-purpose operating systems. The OS only needs to
provide two fairly simple functions, \texttt{\_\_boot} and a scheduler.
This makes the model checker very flexible and easily adaptable to new
needs, besides verification of C and C++ programs. Many modelling
languages, including DVE, ProMeLa or Simulink, can be easily translated
into C code, and with addition of a suitable miniature OS, can be
verified by using our VM approach.

Finally, the virtual machine can remain comparatively simple, which is
very important from reliability standpoint: bugs in the virtual machine
will quietly cause incorrect verification results. However, the OS is
executed by the virtual machine and subject to the same strict checking
that is applied to the user program. Problems caused by the OS will
therefore be detected by the virtual machine and reported to the user,
reducing the risk of falsely positive verification result.

\section{Related Work}\label{related-work}

The idea to use a virtual machine for model checking is not new -- it is
the natural middle ground between compiling the model all the way to a
natively executable successor function (as in
SPIN~\citep{holzmann97:spin} and many later explicit-state model
checkers) and a fully interpreted system (like
UPPAAL~\citep{larsen97:uppaal.nutshel} or earlier versions of
\divine{}~\citep{barnat13:divine}).

One can obtain a suitable virtual machine for use with model checking in
two basic ways: either choose (and possibly adapt) an existing VM with
existing infrastructure (compilers, debuggers, etc.) or design a new one
and re-target an existing compiler, or even create a new one. Of the
newly-designed virtual machines (and their corresponding instruction
sets), the most notable is NIPS VM~\citep{weber10:embedd.virtual}. Along
with other similar designs that emerged from the model checking
community, its focus is on fixed-layout, explicitly finite-state
processes enriched with non-determinism and
synchronisation/communication. VMs of this type are more suitable for
specification-level verification and for verification of embedded
software where use of dynamic memory and dynamic structures in general
is limited.

Successful adaptations of pre-existing instruction sets for use in model
checkers include the Java
PathFinder~\citep{visser03:model.checkin.program} and the \llvm{}-based
\divine{}\,3 and later. The main difference between the two is that the JVM
(the virtual machine underlying Java) is a memory-safe architecture,
while the \llvm{} instruction set is designed primarily as a target for
compilation of unsafe languages such as C and C++. Our current effort is
an evolution of the design used in \divine{}\,3, with emphasis on
separation of concerns and a clean interface. We were able to move large
amounts of code from the virtual machine proper into the (virtualised)
OS, as a consequence of the improved set of primitives (hypercalls)
provided by the virtual machine.

\subsection{Language-Neutral Model
Checking}\label{language-neutral-model-checking}

Many model checkers provide some degree of interoperability with
multiple specification languages. Those efforts are related to \divm{} in
the sense that \divm{} can also be thought of as an interoperability
framework. In explicit-state model checkers, the lowest common
denominator is the functions for enumerating the state space:
\texttt{initial} and \texttt{successors}. At this level, it is usually
quite easy to connect existing unrelated model checking tools: for
example, take the \texttt{successors} function of the Mur\(φ\) model
checker and use \divine{}\,3 to explore the state space. \divine{}\,2 and
\divine{}\,3 both provided interfaces at this level. A more powerful (but
also more complex) alternative is the PINS~\citep{blom09:bridgin.gap}
interface provided by the LTSmin~\citep{kant15:ltsmin} model checker.
The idea behind PINS is to partition the \texttt{successors} function
based on transition groups (where some groups can be, for example,
entirely independent of each other). This additional semantic
information is exposed by the PINS interface, allowing state space
reductions and more efficient search strategies to be implemented in the
host model checker. Nonetheless, static analyses mostly remain specific
to the particular specification language. An additional downside of the
PINS method is that it relies on the state representation being
relatively static. This makes PINS inconvenient to use with extremely
dynamic languages, like those typically used in software development.

In contrast, the \divm{} language fully embraces the dynamic structure of
program states. The model checker interface of \divm{} is, however, nearly
the simplest possible: obtain the initial and successor states. The only
addition on top of the bare minimum is edge labelling, which can be used
to record and present counterexamples. In the PINS approach, the
additional structure is exposed to the model checker and the model
checker makes use of the facilities provided by the model interpreter or
compiler. With \divm{}, the situation is reversed: the VM exposes its
extended functionality to the SUT instead and maintains a trivial
interface with the model checker. This way, static and semi-static
analyses and transformations can work at the level of \llvm{} intermediate
representation, which is already used by many tools.

\subsection{\llvm{}-Based Model Checking}\label{llvm-based-model-checking}

Besides \divm{}, other approaches to \llvm{}-based verification exist. Our
previous work on \divine{}\,3~\citep{barnat13:divine} is largely subsumed
by the current VM-based approach as presented in this paper.

In~\citep{berg13:model.checkin}, the author presents an extension for
LTSmin based on \llvm{} and PINS, for model checking parallel algorithms
under the \emph{partial store order} memory model. Due to its focus on
verification of algorithms and data structures, system-level software is
not considered and \llvm{} is primarily used as convenient means of
verifying algorithms given in the form of C code.

In addition to explicit-state approaches, symbolic, and in particular
SMT-based, tools that build on the \llvm{} IR exist. A prime example in
this category is LLBMC, which works, essentially, by translating \llvm{}
bitcode into an SMT formula which describes the transition function of
the original \llvm{} program. In this case, neither parallel programs nor
system-level code\footnote{To clarify, LLBMC can be used to verify C
  code that is part of system-level software: in fact, a typical
  use-case for symbolic model checkers is analysis of device drivers
  (where the actual device is modelled as a completely non-deterministic
  black box).} is considered. An additional restriction derives from the
fact that the background SMT theory is decidable, and therefore loops
with unknown bounds must be artificially bounded (i.e., LLBMC is a
\emph{bounded model checker}).

A different tool, VVT~\citep{gunther16:vienna.verific.tool}, extends the
approach of LLBMC in two directions. First, it adds support for
concurrency in the input program. Second, it takes a different approach
to encoding the undecidable \llvm{} program into a decidable background
theory, based on \(k\)-induction (and IC3 in particular). However,
system-level interfaces are not considered in VVT either.

\section{Graph-Organised Memory}\label{sec:graphmem}

In \divm{}, the state of the program consists of memory (which is organised
as a graph) and of control registers (described in
Section~\ref{sec:registers}). The semantics of an instruction of the
\divm{} language is, therefore, described by its effect on the machine's
memory and registers. In this section, we will first describe and
justify the graph encoding, then we will describe the semantics of the
memory-related hypercalls (as listed in Table~\ref{tbl:hypercalls}) and
finally, we will discuss the finer details and consequences of this
approach.

A traditional computer treats memory as an array of bytes. Instructions
exist to read data from and store data to a given memory location by
using simple integer indices. Pointers (pieces of data that describe a
location in memory) are, from the point of view of the CPU, really just
integer values. That is, the machine language is untyped and a pointer
can be added to, multiplied or divided just like any other number. Due
to memory virtualisation in basically every modern CPU, which indices
are valid is not determined by the size of the physical memory (as one
would expect if the available memory locations were numbered from 1 to
some \(n\)), but are allocated to a given process by the OS.

Moreover, practically no programs (other than certain parts of operating
system kernels) directly use a flat memory space. Instead, they usually
take advantage of a higher-level interface for management of dynamic
memory, based on the \texttt{malloc} C function. The \texttt{malloc}
function takes care of obtaining unstructured memory from the OS and
divides it in into chunks which can be requested by the program on an
as-needed basis.

\begin{figure}
\centering
\includegraphics[width=1.00000\textwidth]{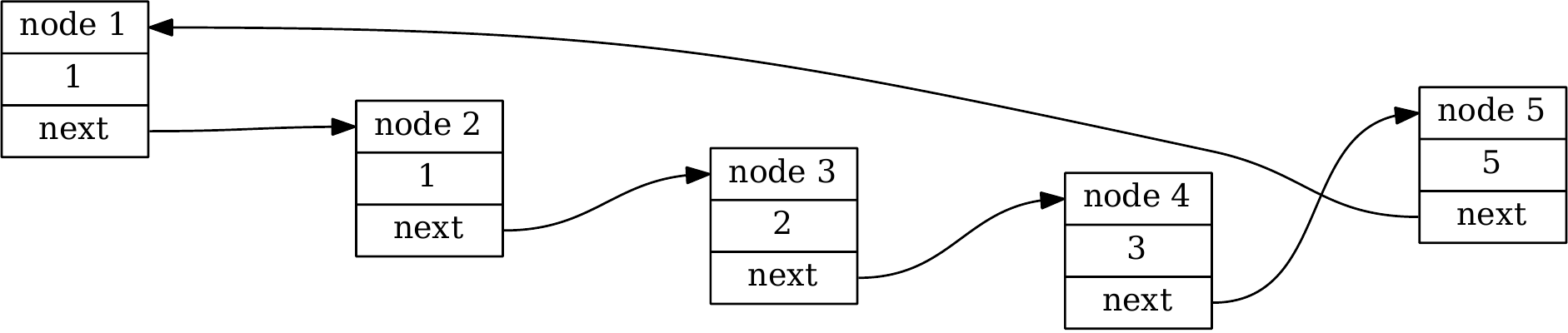}
\caption{A circular linked list, an example of a common data structure
which is often embedded in a heap, taking advantage of
\texttt{malloc}-style memory management.}\label{fig:heap}
\end{figure}

The use of \texttt{malloc}-style memory management is so pervasive in
programs that it is sensible to abstract memory at the level of
\texttt{malloc}-managed objects instead of the more universal,
machine-level flat address space. This is especially true for programs
where concurrency can cause the physical layout of the
\texttt{malloc}-managed heap to vary substantially due to thread
interleaving. However, when we treat the memory as a graph, heap
configurations from different interleavings result in identical graphs.
Hereafter, we will not make a distinction between \emph{heap} as
referring to the \texttt{malloc}-managed portion of memory and the graph
memory structure which exists in the virtual machine (even though the
latter is, in some sense, a superset of the former).

The only requirement of \llvm{} with regards to memory representation is
that pointers need to be fixed-width arithmetic types. It is, however,
neither memory safe nor are memory access instructions type safe. This
poses challenges for model checkers in general and for graph-based
memory organisation in particular. Due to this lack of a static type
system, the virtual machine has no choice but to impose a runtime
(dynamic) type system of its own. The very least that the runtime type
system must do is maintain the distinction between pointers and
non-pointers: otherwise, the graph structure of memory cannot be
recovered. Of course, when such a type system is already in place, it
can be used for other purposes, like tracking uninitialised values. We
will discuss this in more detail in Section~\ref{sec:rttm}.

Finally, with currently available data structures, large heaps (that is,
heaps which store a large number of objects, regardless of their size)
are appreciably more expensive to access and compare. For this reason,
static data is kept in a small number of large objects. For example, all
constant data is kept in a single object, and so is the program code
(\emph{text} in traditional UNIX terminology) and static data (global
variables). Since all the relevant pieces (text, constants, static and
dynamic memory) are part of the heap, they are accessed and represented
uniformly by the virtual machine.

\subsection{Memory Management
Hypercalls}\label{memory-management-hypercalls}

Since \llvm{} bitcode is not tied to a memory representation, its apparatus
for memory management is quite limited. Just like in C, \texttt{malloc},
\texttt{free}, and related functions are provided by libraries, but
ultimately based on some lower-level mechanism, like, for example, the
\texttt{mmap} system call. This is often the case in POSIX systems
targeting machines with a flat-addressed virtual memory system:
\texttt{mmap} is tailored to allocate comparatively large, contiguous
chunks of memory (the requested size must be an integer multiple of
hardware page size) and management of individual objects is done
entirely in user-level code. Lack of any per-object protections is also
a source of many common programming errors, which are often hard to
detect and debug.

It is therefore highly desirable that a single object obtained from
\texttt{malloc} corresponds to a single VM-managed and properly isolated
object. This way, object boundaries can easily be enforced by the model
checker, and any violations reported back to the user. This means that,
instead of subdividing memory obtained from \texttt{mmap}, the
\texttt{libc} running in \divm{} uses \texttt{obj\_make} to create a
separate object for each memory allocation. The \texttt{obj\_make}
hypercall obtains the object size as a parameter and writes the address
of the newly created object into the corresponding \llvm{} register (\llvm{}
registers are stored in memory, and therefore participate in the graph
structure; this is described in more detail in
Section~\ref{sec:frames}). Therefore, the newly created object is
immediately and atomically connected to the rest of the memory graph.

The standard counterpart to \texttt{malloc} is \texttt{free}, which
returns memory, which is no longer needed by the program, into the pool
used by \texttt{malloc}. Again, in \divm{}, there is a hypercall --
\texttt{obj\_free} -- with a role similar to that of standard
\texttt{free}. In particular, \texttt{obj\_free} takes a pointer as an
argument, and marks the corresponding object as \emph{invalid}. Any
further access to this object is a \emph{fault} (faults are described in
more detail in Section~\ref{sec:faults}). The remaining hypercalls in
the \texttt{obj\_} family exist to simplify bookkeeping and are not
particularly important to the semantics of the language.

\subsection{Runtime-Typed Memory}\label{sec:rttm}

Even when the VM has a complete knowledge of the objects residing in
program memory, which can be derived through the API described above,
this alone is not enough to reconstruct the graph structure. The other
necessary component is the knowledge of all pointers stored in the
objects.

At first sight, it may seem that the static type system used by the SSA
portion of \llvm{} (which is easily enforced) could be used to recover
pointer information. The memory portion (that is, non-SSA), however, is
completely untyped, and as such makes it trivial for a program to
circumvent any protection afforded by the type system. The existence of
type casting instructions therefore does not weaken the type system any
further. Since recovering type information statically is very hard and
often quite imprecise, a runtime type system is the only viable
solution. Of course, this does not preclude the use of static analysis
to improve evaluation efficiency.

Fortunately, in a virtual machine, it is easy enough to track type
information through any and all operations performed by the program. The
only limitation is that offsets within a single object should remain
unaffected by the addition of type information. As described
in~\citep{rockai13:improv.state}, the solution to this problem is to
store the type information in a shadow image of the entire address
space. The model checker can keep, in addition to the byte array visible
to the SUT, additional memory associated with each object, in such a way
that this additional (shadow) memory can be easily looked up.

In our implementation of the abstract VM proposed in this paper, we also
use the type system to track whether a particular byte of memory is
\emph{defined}, that is, whether a value has been stored at this
address. The main motivation is that with this information, the model
checker can report suspicious and probably unintended uses of such
undefined values. Due to their low-level nature and focus on execution
speed, both C and C++ elide initialisation code whenever possible. This
elision is, however, not foolproof and can easily lead to unintended
consequences: in some cases, compilers can spot this and emit a warning.
In others, they cannot. The virtual machine can, however, detect
inappropriate uses on all the paths that it explores.

\subsection{Pointer Representation}\label{pointer-representation}

The virtual machine mandates that pointers are represented as tuples,
where the object identifier is separate from the offset within the
object. In our implementation, this is achieved by splitting the 64 bit
pointer into two 32 bit numbers, which are then treated separately.
Moreover, while not strictly required, our implementation stores the
offset part in the least significant bits of the pointer. This somewhat
simplifies implementation of arithmetic instructions when one of the
operands is a (converted) pointer. A strict requirement, however, is
that when the pointer's offset overflows, the pointer becomes
permanently invalid -- the offset must \emph{not} wrap to 0
independently of the object identifier, since the pointer would become
accidentally valid.

Additionally, our implementation also guarantees that the object
identifiers are stable along an execution path: that is, pointers to a
particular object will always use the same numeric object identifier.
However, we acknowledge that in some circumstances, this limitation may
be impractical (see also Section~\ref{sec:symmetry}) and may be lifted
at the expense of banning certain pointer-value-dependent operations in
the program.

Besides regular \emph{heap} pointers, there are 3 additional pointer
types: \emph{global}, \emph{constant} and \emph{code} pointers. While
all data (as opposed to code) is stored on the heap, it is not the case
that each global variable or each constant would reside in a separate
heap object. The virtual machine instead uses \emph{slot}-based
allocation for these types of data, that is, there is a single heap
object for global variables, another for constant data. A \emph{global}
or a \emph{constant} pointer (distinguished from \emph{heap} pointers by
a 2-bit type tag) refer to slots within the designated \emph{globals}
heap object. Slot boundaries are enforced just like object boundaries.

\begin{figure}
\centering
\includegraphics{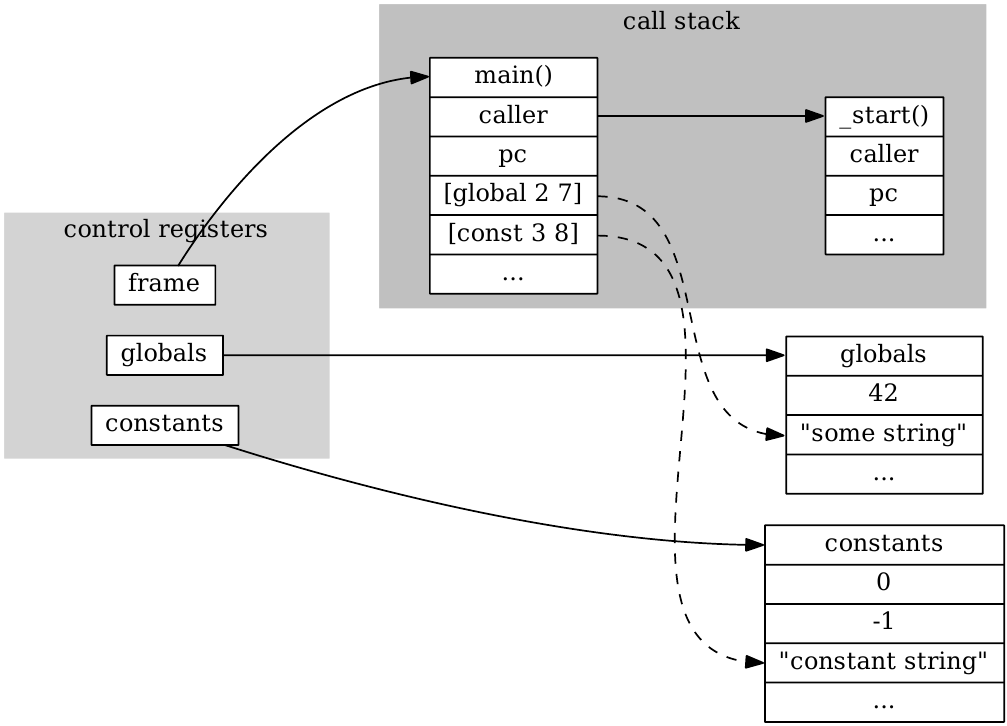}
\caption{Different pointer types. Dashed lines represent indirect
relationships: the value of the respective register is used when
dereferencing such indirect pointers. The first number in an indirect
pointer identifies the slot, the second the offset within the given
slot.}\label{fig:globals}
\end{figure}

The distinction between heap pointers and other pointer types is
important when the OS wishes to implement \texttt{fork()}-like
semantics: with slot-based global variables, different processes can
share the same code (and constants). The OS can set a \emph{control
register} (see also Section~\ref{sec:registers}) to tell the virtual
machine which heap object currently holds global variables. The
situation is illustrated in Figure~\ref{fig:globals}.

\subsection{Memory Protection}\label{memory-protection}

The VM we propose does not have a traditional, page-based MMU (Memory
Management Unit). Nonetheless, since the execution is strictly
controlled, there is a different mechanism which can be employed to
enforce address space separation: if a particular process does not
posses a pointer to a given object, this object cannot be accessed. This
is because the virtual machine enforces object boundaries, therefore, it
is impossible to construct a valid pointer by overflowing a pointer to a
different object (when the offset part of the pointer overflows, the
pointer becomes invalid; likewise, if any operation changes the numeric
value of the object identifier, the value ceases to be a valid pointer).
The only way to access a particular object is, therefore, by obtaining a
pointer to this object, which can be easily prevented by the OS.

\begin{figure}
\centering
\includegraphics[width=1.00000\textwidth]{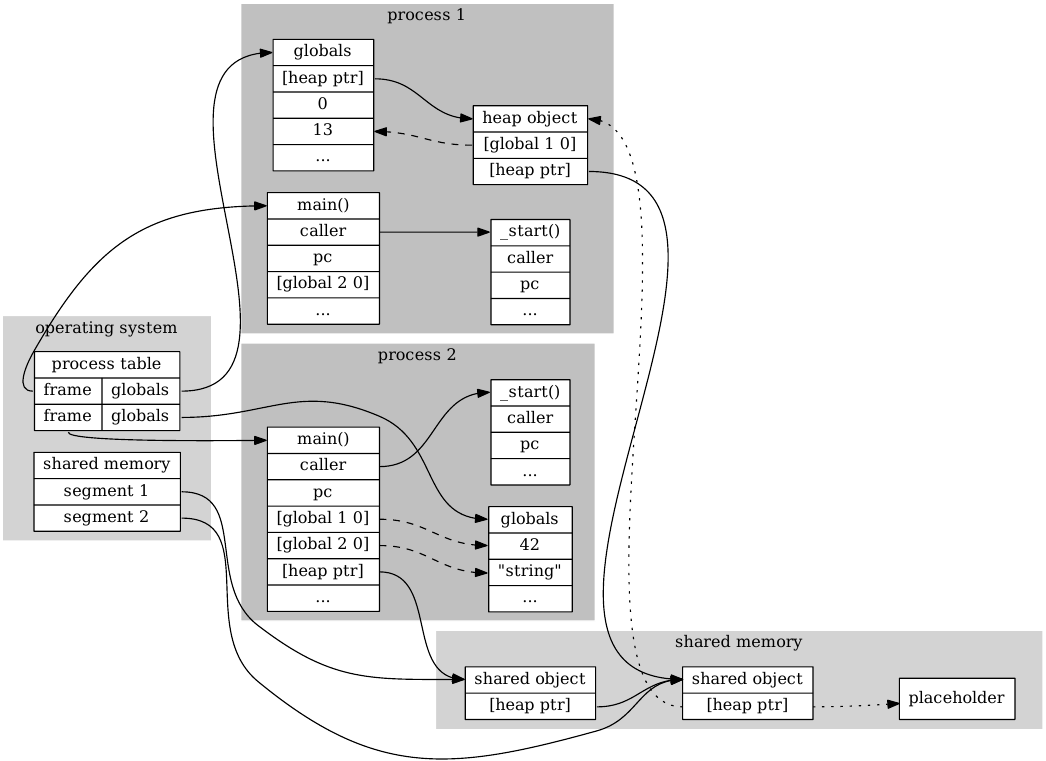}
\caption{Example heap with OS memory, 2 processes and 2 shared memory
segments. Dashed arrows represent \emph{indirect} pointers (see also
Figure~\ref{fig:globals}). The dotted arrow represents a hazardous
pointer (violating memory protection constraints) -- this pointer must
be flipped between the two possible values when processes are switched.
When process 1 executes, it should point into the memory of process~1,
otherwise it should point to the (shared) placeholder
object.}\label{fig:shared}
\end{figure}

The only pitfall of this approach is in the implementation of
inter-process communication (IPC). That is, the enforcement of memory
protection depends on the ability of the OS to invalidate pointers which
are sent to other processes via IPC. If the OS wishes to preserve
pointers that are sent through IPC to another process and then returned
the same way while also enforcing process isolation, it must provide a
translation mechanism. Similar caveats apply to shared memory segments
which may contain pointers to themselves, or to other such segments.
This scenario is illustrated in Figure~\ref{fig:shared}.

\section{Control Flow}\label{control-flow}

In addition to the standard array of features related to control flow
(which are directly inherited from \llvm{}), our virtual machine also needs
to provide features that are more-or-less specific to verification
environments. These include a tightly-controlled scheduling policy,
non-deterministic choices and explicit atomic sections. Additionally,
when compared to a standard, execution-focused VM, there are differences
in how activation frames (that is, the call stack) are represented, and
there are specifics pertaining to control registers.

\subsection{Machine Control Registers}\label{sec:registers}

In addition to the (structured) memory, the virtual machine maintains a
set of \emph{control registers}. Together, these form the entirety of
the execution state of the machine (in other words, the effect of any
given instruction is entirely determined by these two components). The
registers can be read and manipulated through a single hypercall,
\texttt{control}, the interface of which is documented in more detail in
our technical documentation~\citep{rockai16:divine.manual}.

The important distinction between the heap and the registers is that
registers are not part of the \emph{persistent} state of the program:
their values are not taken into account when comparing or storing
program states. They do, however, influence the execution within a
single state space transition (and after evaluation of a given
transition is finished, the values in those registers are cleared).

\subsection{Activation Frames}\label{sec:frames}

Unlike traditional hardware-based implementations of C, our VM does not
use a continuous stack. The present virtual machine takes the approach
of \divine{}\,3 one step further: the execution stack is no longer a
special structure maintained by the model checker itself, but instead is
entirely allocated in the graph-based memory, as a linked list of
activation frames. These frames are fixed in size, as is common when
interpreting \llvm{} bitcode, since they only contain statically-allocated
registers, not variable-sized objects. The latter are always allocated
through the \texttt{alloca} \llvm{} instruction, which in our virtual
machine obtains an appropriately-sized memory object from
\texttt{obj\_make}. Besides \llvm{} registers, the frame contains a pointer
to the caller frame (forming the linked-list structure of the stack) and
a slot for storing the value of the program counter across calls.

Frames are automatically allocated by \texttt{call} and \texttt{invoke}
instructions, but can also be constructed and populated ``manually'' by
the OS when needed. Likewise, the \texttt{ret} instruction deallocates
the current frame, along with all its \texttt{alloca}-obtained memory.

There are 2 main advantages in this stack representation. First, it
means that all the required bookkeeping is done by the graph memory
subsystem (frames are not special in this regard). Second, this
interface naturally allows a high degree of introspection in the SUT.
The OS can, for example, construct an activation frame for the
\texttt{main()} function by using the existing \texttt{make\_obj}
hypercall, instead of requiring additional functionality from the VM.

\begin{figure}
\centering
\includegraphics[width=1.00000\textwidth]{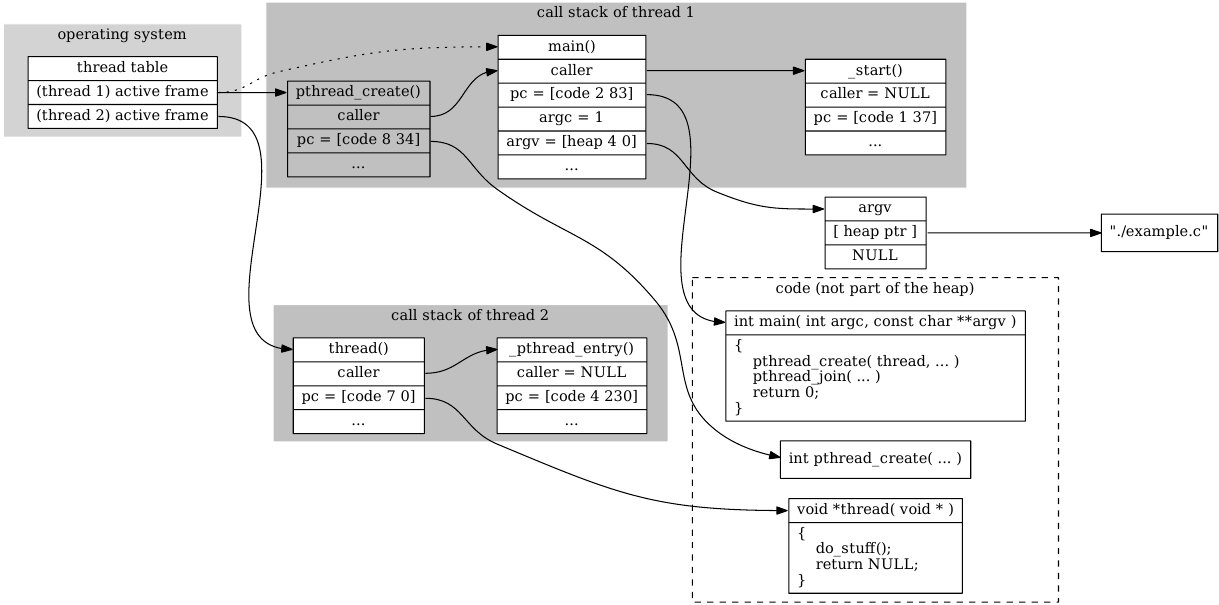}
\caption{Interaction of activation frames and the operating system
scheduler. In this snapshot, \texttt{pthread\_create} is about to
return; as soon as an interrupt happens, the OS will update the active
frame of thread 1 to point to \texttt{main}, as shown by the dotted
arrow. The now-orphaned frame is destroyed by the virtual
machine.}\label{fig:activation}
\end{figure}

\subsection{Scheduling}\label{scheduling}

In a traditional software model checker, threads are first-class,
verifier-managed objects. In our virtual machine design, this does not
need to be the case: it is possible for threads to be implemented within
the virtualised OS in terms of the hypercall interface. Like other
design choices in our approach, this simplifies the VM by moving
responsibility into the OS layer, where functionality is easier to
implement and its correctness is less critical.

In particular, the OS running in the virtual machine is responsible for
providing a \emph{scheduler routine} which decides what to execute in
what order, and the virtual machine uses \emph{interrupts} to return
control to the scheduler whenever the user code executes a possibly
\emph{visible action}. Visible actions must be explicitly marked in the
bitcode. Since visible actions are explicit, the exact semantics of what
is or is not visible is not part of the VM interface: for the VM, an
explicit hypercall, either \texttt{interrupt\_cfl} or
\texttt{interrupt\_mem} is the definition of a visible action.\footnote{In
  a realistic implementation, these explicit \emph{interrupt points} are
  inserted automatically by the bitcode loader in suitable locations.
  For a formalism with shared memory, accesses to memory locations that
  may be shared would constitute visible \emph{memory} actions.
  Likewise, a formalism where invariant loops are possible, all loops
  that may be invariant need to contain a visible \emph{control flow}
  action.} The difference between those two interrupt types is described
in Section~\ref{sec:taubased}.

The variables local to the \emph{scheduler routine} or any functions it
calls are not retained across multiple entries into the scheduler.
Moreover, the scheduler cannot access global variables either. Besides
the transient local variables, all its state must be stored in
explicitly allocated heap objects. One such object is called the
\emph{scheduler state} and a pointer to this object is stored in one of
the control registers. The scheduler can therefore read the value of
this register to access and modify its internal data structures. Other
than the limitations mentioned above, the OS is entirely free to
organise the state information any way it likes.

From the point of view of the state space that is being constructed, the
scheduler decides what the successors of a given state are. When the
verifier needs to obtain successors to a particular state, it executes
the scheduler in that state; the scheduler decides which thread to run
(usually with the help of the non-deterministic choice operator, see
Section~\ref{sec:nondeterminism}) and transfers control to that thread,
by instructing the virtual machine to execute a particular activation
frame (a pointer to which is stored for this purpose in the scheduler
state).

\subsection{Atomic Sections}\label{atomic-sections}

In the present design, support for explicit atomic sections in the
virtual machine is not strictly necessary. Since the virtual machine
supports standard atomic memory access instructions, it is possible to
implement mutual exclusion on top of these. However, this is
inefficient: it is, in general, impossible to recover the relationship
between atomic operations or explicit locks and the memory accesses they
guard (this is one of the reasons we need a model checker in the first
place). This would make a system that does not make use of atomic
sections substantially less efficient when running in the virtual
machine, due to a large number of extra interrupts.

Additionally, when static analysis can prove that a particular section
of code is protected by a mutual exclusion device (that is, all relevant
memory locations it accesses), it can insert an explicit atomic section,
making subsequent verification more efficient. Likewise, this ability of
the virtual machine can be used to implement adaptive-precision model
checking, where certain operations are assumed to be thread-safe, again
making the verification process less demanding.

\subsection{Non-deterministic Choice and
Counterexamples}\label{sec:nondeterminism}

It is often the case that the behaviour of a program depends on outside
influences, which cannot be reasonably described in a deterministic
fashion and wired into the SUT. Such influences are collectively known
as the \emph{environment}, and the effects of the environment translate
into non-deterministic behaviour. A major source of this non-determinism
is thread interleaving -- or, equivalently, the choice of which thread
should run next after an interrupt.

In our design, all non-determinism in the program (and the operating
system) is derived from uses of the \texttt{choose} hypercall (which
non-deterministically returns an integer between 0 and a given number).
Since everything else in the SUT is completely deterministic, the
succession of values produced by calls to \texttt{choose} specifies an
execution trace unambiguously. This trait makes it quite simple to store
counterexamples and other traces in a tool-neutral, machine-readable
fashion. Additionally, hints about which interrupts fired can be
included in case the counterexample consumer does not wish to reproduce
the exact interrupt semantics of the given VM implementation.

Finally, the \texttt{trace} hypercall serves to attach additional
information to transitions in the execution graph. In particular, this
information then becomes part of the counterexample when it is presented
to the user. For example, the \texttt{libc} provided by \divine{} uses the
\texttt{trace} hypercall in the implementation of standard IO functions.
This way, if a program prints something to its standard output during
the violating run, this output becomes visible in the counterexample.

\section{Property Specification}\label{property-specification}

An important aspect of a verifier is the specification of desirable
properties of the program. In our design, this task is largely delegated
to the OS. However, there are 2 aspects of property specification that
require support from the VM. First, there are many circumstances in
which the VM can detect problematic behaviour in the program that would
be impractical to detect by other means. This includes out-of-bounds
memory accesses, use of undefined values, mismatches between formal and
actual arguments in \texttt{call} instructions and so on. For maximal
flexibility, these conditions are not directly exposed as program
properties, but are instead signalled to the OS by invoking a
\emph{fault handler}. This fault handler is then free to decide how to
respond to this particular fault and whether to signal a property
violation or not.

The other area where the virtual machine must be involved is the
communication of the operating system with the verification algorithm.
That is, the OS must be able to signal the fact that a particular
transition is an error transition (or an accepting transition, in case
of \(\omega\)-regular properties). For this purpose, a pair of bits is
reserved in one of the machine state registers, corresponding to either
an error or an accepting transition. In turn, the verifier obtains this
information from the VM to inform its decisions.

\subsection{Faults}\label{sec:faults}

When the program attempts to execute an illegal instruction, the virtual
machine will enter a designated \emph{fault handler} instead of
continuing execution. The reasons why the instruction is deemed illegal
are various, but they roughly correspond to conditions checked by
standard CPUs, which on POSIX systems translate to signals. The checks
done by the VM are, however, stricter than the corresponding checks in
normal execution environments. This includes more granular information
about objects, impossibility to overflow a pointer from one object into
another, tracking of undefined values, checks on correct use of variadic
function arguments, immutability of constant data, validity of target
addresses in branch instructions and strict checking of validity of heap
operations. Additionally, all uses of the hypercall interface are
strictly checked for conformance with the specification.

Faults are, in principle, not fatal: the fault handler may choose to
continue execution despite the raised error. For this reason, the VM
passes a \emph{continuation}\footnote{This consists of the code pointer
  and the frame where execution would continue if the instruction
  succeeded.} to the fault handler, which it may choose to invoke;
alternatively, the fault handler may abort execution and report the
error to the verifier, and through that, to the user. This mechanism is
especially important when a fault arises due to a control flow
instruction -- typically, the target of a conditional branch instruction
could depend on an undefined value. In this case, the continuation is
chosen \emph{as if} the value was defined and had the specific value
observed at the point of the fault.\footnote{All undefined values come
  into existence with \texttt{0} as their \emph{as if defined} value,
  but may be combined with other (defined) values to obtain
  \emph{undefined} but non-zero values. For example, for
  \texttt{int\ a,\ b\ =\ a\ +\ 7} the value of \texttt{b} would be
  \emph{undefined} \texttt{7}.}

\subsection{Monitors and LTL}\label{monitors-and-ltl}

Thanks to the flexible scheduler design, it is very easy to implement
properties as \emph{monitors}, that is, additional finite-state automata
which synchronise with the executing program to observe its behaviour.
These automata can then either flag error transitions (for safety
verification) or mark accepting transitions (for liveness verification).
Classical algorithms for automata-based LTL model checking can then be
used to verify LTL properties translated into monitors.

\section{Reduction and Abstraction}\label{reduction-and-abstraction}

In our earlier work~\citep{rockai13:improv.state}, we described a number
of reduction techniques tailored toward verification of \llvm{} bitcode.
Many of these can be recovered in the new VM-based approach without
significant semantic changes. However, some of the technical solutions
are rather different. In the case of \(τ+\)reduction, we have opted to
insert interruption points (instructions) explicitly into the program,
instead of co-opting memory access instructions for this purpose. The
advantages are two-fold: it substantially simplifies tools which work
with counterexamples, since they do not need to know anything at all
about τ reduction -- the trace from the verifier can, without
significant expense, include the information whether a given interrupt
fired or did not fire. This way, replaying the counterexample involves
simplified implementations of both \texttt{choose} and of the
\texttt{interrupt\_*} family of hypercalls, where they simply read the
next entry in the counterexample trace to either obtain the return value
or to decide whether an interrupt should fire.

\subsection{τ-Based Reductions}\label{sec:taubased}

Within the verifier, the implementation of \texttt{interrupt\_mem} and
\texttt{interrupt\_cfl} are where most of the τ reduction logic is
implemented. The \texttt{interrupt\_mem} hypercall signals to the VM
that a memory operation is about to be executed, along with the affected
address and type of memory access. On the other hand,
\texttt{interrupt\_cfl} signals that a loop in the program state space
may have formed. In a simple implementation, both these hypercalls could
simply cause an unconditional interrupt, without compromising
correctness in any way. The additional information (the type of
interrupt -- \texttt{cfl} vs \texttt{mem} -- and the memory location and
access type in the latter) is provided in order to improve efficiency.
Clearly, if an interrupt can be safely suppressed, fewer distinct
program states need to be stored, saving both time and space.

In our current version, the control flow interrupts are treated as
described in~\citep{rockai13:improv.state}, that is, the VM keeps track
of program counter values that execution passed through, and only causes
an interrupt if the particular instruction was already evaluated once
within the given state space transition. In addition
to~\citep{rockai13:improv.state}, the new implementation only stores
program counter values that correspond to \texttt{interrupt\_cfl} calls,
reducing evaluation overhead for other instructions.

Likewise, memory interrupts can often be suppressed: first, multiple
independent loads can be all coalesced until a store instruction is
encountered, or until a load from an address that was already used is
repeated~\citep{still16:llvm.transf}. That is, an interrupt is only
performed for \texttt{store}-type instructions, or for repeated
\texttt{load} instructions.

Second, some stores and some repeated loads are also invisible; in
particular, when a memory location is only reachable from a single
thread, all interrupts related to that location can be
suppressed~\citep{rockai13:improv.state}. We say that a memory object is
\emph{thread private} when no other thread is in possession of a pointer
to this object. Since only one thread can access the memory, changes in
this memory cannot be observed by any other thread. Since the VM
maintains the entire memory as an oriented graph, a simple heuristic can
be used to suppress interrupts related to such non-observable memory
operations. In particular, the VM can maintain a set of \emph{shared}
objects -- those that are not \emph{thread private}. The invariant
property of the set of \emph{shared} objects is that it is closed under
reachability along pointers (edges of the memory graph). When the
program starts, global variables (which are accessible from any thread)
are initially included in this set. Likewise, when a new thread is
created and is given access to some objects, those objects are included
in the set. All other operations simply maintain the invariant: when a
pointer to object \(A\) is written to \(B\) and \(B\) is shared, all
objects reachable from \(A\) (including \(A\) itself) are added to the
\emph{shared} set.

Since the VM has no concept of threads, it is the responsibility of the
OS to inform the VM when new objects become shared via thread creation.
That is, when a pointer to a previously private object is written
directly to another private object, owned by a different thread, the
operating system must call the \texttt{obj\_shared} hypercall on this
pointer. Outside of the operating system, the only way to share new
objects is by writing their addresses into an already shared memory
location.

\subsection{Symmetry-Based Reductions}\label{sec:symmetry}

Like equivalent thread interleavings in τ reductions, heap symmetry is
an important source of redundancy in the state space. Since the virtual
machine has access to the graph structure of memory, it can easily
compute a canonic form for comparison purposes. One simple approach is
to execute DFS from the root object (that is, the object corresponding
to the \emph{state} of the scheduler) and sequentially assign numbers to
objects in pre-order, adjusting pointers along the way. Another is to
use a mark-and-copy garbage collector to compact the entire memory into
a contiguous chunk and store this chunk in a hash table -- this is
basically the approach \divine{}\,3 uses. Both these approaches have an
important problem though: the meaning of values derived from
pointer-to-number conversions and the pointer ordering are not preserved
during execution. In some cases, like hash tables with pointer-based
keys, this can cause incorrect results. Therefore, the recommended way
to implement heap symmetry reduction is to only use the canonic form for
comparison purposes, but for successor generation, store a particular
non-canonic form. This way, continuity of pointer-derived values can be
guaranteed along any given execution. Of course, more sophisticated --
and more efficient -- approaches based on partial hashes are possible.

Additionally, since all persistent data in the program are now stored
uniformly in the graph structure, the benefits of symmetry reduction
also extend to stacks, global variables and other auxiliary data
structures. This effect therefore also makes it possible to avoid
exploring states where multiple instances of the same thread only differ
in order of execution among themselves.

\subsection{Abstractions and Symbolic Data}\label{sec:symbolic}

In addition to compatibility with important state space reductions, the
proposed virtual machine works seamlessly with transformation-based
abstractions~\citep{rockai15:model.checkin.softwar}. While in theory,
all the environment-induced non-determinism is the same, reading a
number from the environment causes non-deterministic branching of a very
high degree (corresponding to the number of distinct values that can be
represented by a given data type, say \(2^{32}\) for a typical
\texttt{int} value). This is clearly impractical. For this reason, it is
important that our virtual machine can be co-opted for abstraction-based
model checking. Since the method described
in~\citep{rockai15:model.checkin.softwar} works by transforming code
ahead of time, there are only two requirements on the virtual machine:
first, it needs to support non-deterministic choice, since abstracted
operations could have indeterminate results; second, it must be able to
provide machine-readable counterexamples (ideally in a form that is easy
to process). Both these requirements are easily fulfilled in the
proposed design (see also Section~\ref{sec:nondeterminism}).

Finally, formula- or decision-diagram-based symbolic data can be
represented as a type of abstract domain, and as such is subsumed by the
above. The difference is that when symbolic data is used, this must be
reflected in the decision procedure -- at minimum, state comparison must
be altered to use semantic formula equivalence on the symbolic portion
of the state, instead of structural comparison used for
explicitly-represented portions of memory. This translates to an
additional requirement for the virtual machine, that is, the interface
with the verification core needs to support a sufficiently simple method
to read and interpret the heap. However, this is a purely technical
problem: nothing in the semantics of the VM prevents such interface, and
our reference implementation in \divine{}\,4 does provide this access.

\section{Implementation \& Evaluation}\label{implementation-evaluation}

Besides providing the specification of the interface and (informal)
semantics of the virtual machine, we also make available the source code
of a reference implementation.\footnote{Instructions for downloading the
  source code can be found at
  \url{http://divine.fi.muni.cz/download.html}. The code is covered by
  the ISC (simplified BSD) open-source licence.} While the \divm{} language
is, in principle, based on the \llvm{} instruction set and therefore our
implementation relies on \llvm{} for C and C++ compiler frontends and for
code transformation, in principle, the hypercall interface could be
adapted to other instruction sets. This is because it is fully possible
to realize the hypercall interface as C functions, and as such, it could
be combined with a different instruction set and used from any
C-compatible programming language. In addition to the VM itself, we
provide a C++ implementation of a small, verification-focused operating
system, \dios{}.\footnote{Available from the same source repository.} The
main focus of \dios{} is to support verification of C and C++ programs
written using POSIX APIs.

There is an additional implementation-related benefit of \divm{}. Namely,
the virtual machine itself does not depend on \llvm{} libraries. Since \llvm{}
does not provide a stable interface and constitutes a substantial
dependency, not linking to \llvm{} makes the resulting code more portable
and easier to build. The compiler and transformation passes of course
still require the \llvm{} infrastructure, but these can be kept separate
from the model checking tool itself.

\subsection{Benchmarks}\label{sec:benchmarks}

To evaluate the work presented in this paper, we have used a set of 1045
benchmarks -- each of is a C or a C++ program. Out of those programs,
the majority (926) is correct, while 119 contain an error. Most of the
programs are C++, with the exception of the ``svc-pthread'', ``pt-w32''
and ``libc-std'' categories, which are written in standard C (and they
make use of POSIX threads, outside of the ``libc-std'' category). The
``alg'' category includes sequential algorithmic and data structure
benchmarks, the ``courses'' category contains unit tests for student
assignments in various C++ courses, including concurrent data structures
and other parallel programs, ``libcxx'' contains a selection of the
\texttt{libc++} testsuite, ``bricks'' contains unit tests for various
C++ helper classes, including concurrent data structures, ``llvm-bench''
category contains programs from the \llvm{} test-suite and the
``svc-pthread'' category includes pthread-based C programs from the
SV-COMP benchmark set. The ``libc-std'' category contains tests of
\texttt{libc} functionality, while \texttt{pt-w32} test the POSIX
threading API. The ``other'' category is a selection of programs which
did not fit any other category.

In most of the programs, it was assumed that \texttt{malloc} and
\texttt{new} never fail, with the notable exception of part of the
``bricks'' category unit tests.

\subsection{Results}\label{results}

We have executed all the benchmarks described in
Section~\ref{sec:benchmarks} with 4 tools. The approach of the present
paper is represented by \divine{}\,4, an explicit-state model checker based
on \divm{}. Our primary comparison was with \divine{}\,3, which is an earlier
version of this tool, in some sense a predecessor to our current
approach based on \divm{}. Two variants of \divine{} 3 were used, because in
the course of evaluation, it was discovered that \divine{} 3 in its
original version suppresses certain valid thread interleavings. Since
this omission did not lead to any false negatives on the benchmark set,
we include both the original results (with the interleaving incorrectly
suppressed) and a fixed version (marked as ``D3+p'' in the comparison
tables).

The results are very promising: we did not see any substantial
regression caused by the higher abstraction level and increased
isolation of components in \divine{} 4. Quite to the contrary, in many
cases, the new approach is substantially more efficient, which we
ascribe to better isolation of components: smaller and simpler
components are usually easier to optimise than large, complex ones.

The verification time and state count for \divine{} 4 for all models are
summarised in Table~\ref{tbl:D4}. This is the baseline to which all
other tools are compared.

\hypertarget{tbl:D4}{}
\begin{longtable}[]{@{}lrrr@{}}
\caption{\label{tbl:D4}Benchmark results for \divine{} 4, an explicit-state
model checker based on \divm{}. }\tabularnewline
\toprule
tag & models & D4 search & D4 states\tabularnewline
\midrule
\endfirsthead
\toprule
tag & models & D4 search & D4 states\tabularnewline
\midrule
\endhead
bricks & 295 & 3:07:18 & 7233\,k\tabularnewline
courses & 28 & 33:30 & 5399\,k\tabularnewline
libcxx & 461 & 42:37 & 2182\,k\tabularnewline
libc-std & 81 & 26:53 & 3787\,k\tabularnewline
pt-w32 & 10 & 22:30 & 1680\,k\tabularnewline
llvm-bench & 22 & 2:42:44 & 10.7\,M\tabularnewline
svc-pthread & 17 & 15:40 & 1685\,k\tabularnewline
other & 12 & 4:29 & 298.8\,k\tabularnewline
\textbf{total} & 926 & 8:15:44 & 33.0\,M\tabularnewline
\bottomrule
\end{longtable}

\subsection{Comparison to \divine{} 3}\label{comparison-to-divine-3}

The previous version of \divine{} is based on a custom \llvm{} bitcode
interpreter, with an ad-hoc set of extensions. Unlike \divm{}, it has a
built-in notion of threads based on asynchronous execution and special
operations for thread creation and management, on which a
pthread-compatible API is built. Like \divine{}\,4, it is an explicit-state
model checker and also contains a large subset of standard C and C++
libraries. However, its libraries are less complete, which is part of
the reason more than a half of the models could not be verified with
\divine{}\,3.

Comparison of verification time and the number of states explored is
shown in Table~\ref{tbl:D4D3}. In this case, we can see that \divine{}\,4
is much faster in all benchmark categories with the exception of
``svc-pthread'' and ``pt-w32'' which both focus on threaded programs and
are written in plain C. Additionally, \divine{}\,4 reduced the state spaces
more successfully, with the sole exception of ``pt-w32''.

As outlined above, the difference in performance with thread-heavy
models was tracked down to an omission of a particular set of
interleavings in \divine{}\,3. When this problem is corrected, the time and
state space size difference in ``pt-w32'' is reversed (with 2 models now
running out of memory in \divine{}\,3). The results of this comparison are
shown in Table~\ref{tbl:D4D3p}.

\hypertarget{tbl:D4D3}{}
\begin{longtable}[]{@{}lrrrrr@{}}
\caption{\label{tbl:D4D3}Comparison of \divine{}\,4 and \divine{}\,3. Out of
the 926 error-free models, it was only 457 possible to verify 457 with
\divine{}\,3, typically due to incomplete standard libraries.
}\tabularnewline
\toprule
tag & models & D4 search & D3 search & D4 states & D3
states\tabularnewline
\midrule
\endfirsthead
\toprule
tag & models & D4 search & D3 search & D4 states & D3
states\tabularnewline
\midrule
\endhead
courses & 1 & 0:00 & 0:02 & 57 & 287\tabularnewline
libcxx & 344 & 20:15 & 5:48:12 & 787.4\,k & 4190\,k\tabularnewline
libc-std & 76 & 0:51 & 1:27 & 33.7\,k & 54.1\,k\tabularnewline
llvm-bench & 3 & 2:11 & 27:17 & 306.6\,k & 2351\,k\tabularnewline
pt-w32 & 10 & 22:30 & 6:53 & 1680\,k & 542.7\,k\tabularnewline
svc-pthread & 16 & 15:24 & 27:15 & 1658\,k & 4123\,k\tabularnewline
other & 7 & 4:03 & 9:22 & 263.4\,k & 302.6\,k\tabularnewline
\textbf{total} & 457 & 1:05:17 & 7:00:33 & 4729\,k &
11.6\,M\tabularnewline
\bottomrule
\end{longtable}

\hypertarget{tbl:D4D3p}{}
\begin{longtable}[]{@{}lrrrrr@{}}
\caption{\label{tbl:D4D3p}Comparison of \divine{}\,4 and modified
\divine{}\,3, where additional interleavings were taken into account
(original \divine{}\,3 did not allow newly starting threads to be delayed).
}\tabularnewline
\toprule
tag & models & D4 search & D3+p search & D4 states & D3+p
states\tabularnewline
\midrule
\endfirsthead
\toprule
tag & models & D4 search & D3+p search & D4 states & D3+p
states\tabularnewline
\midrule
\endhead
courses & 1 & 0:00 & 0:02 & 57 & 287\tabularnewline
libcxx & 342 & 17:08 & 4:20:52 & 687.8\,k & 3605\,k\tabularnewline
libc-std & 76 & 0:51 & 1:29 & 33.7\,k & 54.1\,k\tabularnewline
llvm-bench & 2 & 1:31 & 17:18 & 306.5\,k & 808.4\,k\tabularnewline
pt-w32 & 8 & 4:54 & 11:44 & 414.9\,k & 849.3\,k\tabularnewline
svc-pthread & 12 & 0:56 & 0:50 & 113.9\,k & 183.2\,k\tabularnewline
other & 7 & 4:03 & 9:23 & 263.4\,k & 302.6\,k\tabularnewline
\textbf{total} & 448 & 29:27 & 5:01:40 & 1820\,k &
5803\,k\tabularnewline
\bottomrule
\end{longtable}

\subsection{Comparison to ESBMC 4.1}\label{comparison-to-esbmc-4.1}

The other tool we have chosen for comparison is ESBMC, an SMT-based
symbolic model checker with support for C++. Part of the reason for this
choice was that many of our benchmarks are C++, and most tools, even
those based on \llvm{}, can only handle C. Unfortunately, the C++ support
in ESBMC is incomplete, and only supports older C++ revisions (C++98,
but C++11 is already in widespread use). Likewise, the support for
standard C++ library is quite limited in ESBMC, since it was only able
to verify 42 out of over 400 tests in the ``libcxx'' category. Another
major problem we encountered in ESBMC is inefficient support for
threads, which can be seen in the very large time difference in the
``svc-pthread'' category. Overall, it was possible to verify only a
small number of models and the models that could be verified often took
much longer in ESBMC than they did \divine{}\,4.

\begin{longtable}[]{@{}lrrrr@{}}
\caption{Comparison of \divine{}\,4 and ESBMC 4.1.}\tabularnewline
\toprule
tag & models & D4 search & ESBMC search & D4 states\tabularnewline
\midrule
\endfirsthead
\toprule
tag & models & D4 search & ESBMC search & D4 states\tabularnewline
\midrule
\endhead
libcxx & 42 & 0:23 & 0:07 & 217\tabularnewline
libc-std & 6 & 0:03 & 0:03 & 322\tabularnewline
llvm-bench & 5 & 8:18 & 1:56:37 & 1819\,k\tabularnewline
pt-w32 & 1 & 0:00 & 0:00 & 7\tabularnewline
svc-pthread & 4 & 1:35 & 46:43 & 280.8\,k\tabularnewline
other & 2 & 0:01 & 0:08 & 390\tabularnewline
\textbf{total} & 60 & 10:22 & 2:43:39 & 2101\,k\tabularnewline
\bottomrule
\end{longtable}

\section{Conclusion}\label{conclusion}

We have shown that many features expected in an explicit-state model
checker are readily recovered in the proposed virtual-machine-based
approach. In many cases, the solutions are simpler and cleaner and
without a substantial performance penalty, as demonstrated in
experiments. Additionally, both the greater overall simplicity and the
fact that a large portion of verification-related code can be executed
in the strict, error-checking virtual machine, contribute to appreciable
improvements in robustness.

Rather importantly, the proposed virtual machine interface is compatible
with all the important techniques that improve efficiency of
explicit-state model checking -- including reductions based on partial
orders and path compression (τ-reductions) and reductions based on heap
configuration symmetries. Likewise, it can be easily combined with
abstraction techniques based on program transformation.

\bibliographystyle{elsarticle-num}
\bibliography{common}

\end{document}